\documentclass[sigconf,natbib]{acmart}

\usepackage{bm}
\usepackage{booktabs}
\usepackage{url}
\usepackage{hyperref}
\usepackage{multirow}
\usepackage[ruled,vlined,boxed,linesnumbered]{algorithm2e}
\usepackage{caption}
\usepackage{graphicx}
\usepackage{subcaption}
\usepackage{enumitem}
\usepackage{balance}
\usepackage{caption}
\usepackage{float}

\AtBeginDocument{%
  }

\setcopyright{acmlicensed}
\copyrightyear{2018}
\acmYear{2018}
\acmDOI{XXXXXXX.XXXXXXX}
\acmConference[Woodstock '18]{Woodstock '18: ACM Symposium on Neural
  Gaze Detection}{June 03--05, 2018}{Woodstock, NY}
\acmISBN{978-1-4503-XXXX-X/2018/06}

\begin{document}

% \fancyhead{}

%%
%% The "title" command has an optional parameter,
%% allowing the author to define a "short title" to be used in page headers.
\title{NGA: Non-autoregressive Generative Auction with Global Externalities for Advertising Systems}

%%
%% The "author" command and its associated commands are used to define
%% the authors and their affiliations.
%% Of note is the shared affiliation of the first two authors, and the
%% "authornote" and "authornotemark" commands
%% used to denote shared contribution to the research.
% \author{Anonymous}

\author{Zuowu Zheng$^*$, Ze Wang$^*$, Fan Yang$^\dagger$, Wenqing Ye, Weihua Huang, Wenqiang He,\\ Teng Zhang, Xingxing Wang}

\affiliation{%
 \institution{Meituan, Shanghai, China}
 \city{}
 \country{}
}
\email{{zhengzuowu, wangze18, yangfan129, yewenqing, huangweihua02, hewenqiang05}@meituan.com}
\email{{zhangteng09, wangxingxing04}@meituan.com}

%%
%% By default, the full list of authors will be used in the page
%% headers. Often, this list is too long, and will overlap
%% other information printed in the page headers. This command allows
%% the author to define a more concise list
%% of authors' names for this purpose.

% \renewcommand{\shortauthors}{Trovato and Tobin, et al.}
\renewcommand{\shortauthors}{Zheng et al.}

%%
%% The abstract is a short summary of the work to be presented in the
%% article.
\begin{abstract}
\renewcommand{\thefootnote}{\fnsymbol{footnote}}
\footnotetext[1]{Equal contribution.}
\renewcommand{\thefootnote}{\fnsymbol{footnote}}
\footnotetext[2]{Corresponding author.}
\renewcommand{\thefootnote}{\arabic{footnote}}

Online advertising auctions are fundamental to internet commerce, demanding solutions that not only maximize revenue but also ensure incentive compatibility, high-quality user experience, and real-time efficiency. While recent learning-based auction frameworks have improved context modeling by capturing intra-list dependencies among ads, they remain limited in addressing global externalities and often suffer from inefficiencies caused by sequential processing. In this work, we introduce the Non-autoregressive Generative Auction with global externalities (NGA), a novel end-to-end framework designed for industrial online advertising. NGA explicitly models global externalities by jointly capturing the relationships among ads as well as the effects of adjacent organic content. To further enhance efficiency, NGA utilizes a non-autoregressive, constraint-based decoding strategy and a parallel multi-tower evaluator for unified list-wise reward and payment computation. Extensive offline experiments and large-scale online A/B testing on commercial advertising platforms demonstrate that NGA consistently outperforms existing methods in both effectiveness and efficiency.
\end{abstract}

%%
%% The code below is generated by the tool at http://dl.acm.org/ccs.cfm.
%% Please copy and paste the code instead of the example below.
%%
\begin{CCSXML}
<ccs2012>
   <concept>
       <concept_id>10002951.10003260.10003272.10003275</concept_id>
       <concept_desc>Information systems~Display advertising</concept_desc>
       <concept_significance>500</concept_significance>
       </concept>
 </ccs2012>
\end{CCSXML}

\ccsdesc[500]{Information systems~Display advertising}

%%
%% Keywords. The author(s) should pick words that accurately describe
%% the work being presented. Separate the keywords with commas.
\keywords{Generative Auction, Mechanism Design, Externalities}

%%
%% This command processes the author and affiliation and title
%% information and builds the first part of the formatted document.
\maketitle

\section{Introduction}
Online advertising \cite{jansen2008sponsored,liao2022cross,liao2022deep,shi2023pier} has emerged as the internet industry's primary revenue source, offering advertisers a cost-effective and targeted channel to reach millions of users. When a user request arrives, platforms execute real-time auctions to allocate ads across available slots, and determine payments for winning advertisers \cite{liao2022nma,wang2022learning,wang2022hybrid,zhao2021dear,xie2021hierarchical,yan2020ads}. An optimal auction mechanism must satisfy multiple objectives: maximizing platform revenue, maintaining key economic properties, including incentive compatibility (IC) and individual rationality (IR) \cite{myerson1981optimal}, and meeting computational complexity requirements for real-time operation \cite{shi2023mddl, qiu2025one, zheng2025ega}.

Traditionally, the Generalized Second Price (GSP)~\cite{edelman2007internet} auction and its variants like uGSP~\cite{bachrach2014optimising} and DeepGSP~\cite{zhang2021optimizing} have served as the industry standard, which assumes that click-through rate (CTR) is separable and user clicks only rely on the ad itself. Consequently, GSP is inherently limited by its inability to capture contextual dependencies, i.e., externalities~\cite{ghosh2008externalities,gatti2012truthful,hummel2014position}, between ads and their surrounding content.
To address this, recent advances in computation have motivated the development of learning-based auction frameworks~\cite{zhang2021survey}. For example,  Deep Neural Auction (DNA)~\cite{liu2021neural} and Score Weighted VCG (SW-VCG)~\cite{li2023learning} extend classical auctions by incorporating online feedback into end-to-end learning pipelines. However, these models suffer from the evaluation-before-ranking dilemma, where the rank score must be predicted before knowing the final allocated sequence, limiting its capacity to model set-level externalities.
Other approaches attempt to integrate optimality into auction design. MIAA~\cite{li2024deep} presents a deep automated mechanism that integrates ad auction and allocation, which simultaneously decides the ranking, payment, and display position of the ad. However, MIAA is limited to the allocation of a single ad, and its computational complexity grows rapidly when handling a large number of candidate ads.
Contextual Generative Auction (CGA)~\cite{zhu2024contextual} addresses the limitations of traditional and learning-based ad auctions by explicitly modeling permutation-level externalities through an autoregressive allocation model and a gradient-friendly reformulation of incentive compatibility, enabling end-to-end optimization of both allocation and payment. It represent a significant step forward in advancing auction design for online advertising.

However, several critical limitations still remain in existing generative auction frameworks:
\textbf{i)} While approaches like CGA consider dependencies among ads, they typically overlook the broader impact of neighboring organic content (non-sponsored) on user experience and auction performance. In practice, the position of an ad and its context within the page, including adjacent organic items, play a crucial role in determining the outcome of the auction, yet these global externalities remain underexplored.
\textbf{ii)} Current generative auction methods often employ autoregressive generation processes, which build the auction outcome sequentially. Although this allows for some context sensitivity, the sequential nature increases inference latency, presenting scalability and responsiveness challenges, especially in settings with many ad slots or candidate items.
\textbf{iii)} In existing frameworks, the evaluation and payment computation steps are typically carried out in a serial fashion, further exacerbating latency issues in large-scale, real-time environments.

To address the aforementioned limitations in generative auction frameworks, we propose a novel approach, Non-autoregressive Generative Auction with global externalities (NGA). \textit{Firstly}, NGA incorporates accurate modeling of auction externalities by jointly considering both ad candidate dependencies and the influence of surrounding organic content. This enables more precise, context-aware decision making, effectively capturing the real-world impact of ad placements within heterogeneous page environments.
\textit{Secondly}, we design a non-autoregressive and constrained decoding based mechanism that significantly improves decision-making efficiency. \textit{Last but not least}, NGA integrates evaluator and payment computations into a unified framework that supports parallel prediction, significantly reducing overall latency and improving scalability.

Our contributions are as follows:
\begin{itemize}[leftmargin=2em]
    \item We propose a novel generative auction framework NGA that comprehensively models global externalities, which enables more accurate, context-aware auction outcomes and better reflects real-world user experiences.
    \item We design a non-autoregressive and constraint decoding based strategy together with a multi-tower parallel evaluator for reward and payment computation. This integrated approach significantly improves inference efficiency and scalability.
    \item Extensive offline experiments and large-scale online A/B testing on commercial advertising platforms demonstrate that our approach achieves superior performance in both effectiveness and efficiency compared to state-of-the-art baselines.
\end{itemize}

\section{Preliminary}
% \subsection{Task Formulation}
We formalize a typical task of joint ad auction and allocation in online advertising systems. Given a page view (PV) request from user $u$, there are $n$ candidate ads and $m$ candidate organic items. The organic sequence is assumed to be pre-ranked by an upstream module based on estimated order volume, and its internal order will remain fixed\footnote{This reflects common platform design constraints where organic content is ranked independently and must preserve user experience.}.
The system selects a final ranked list for $k$ slots ($k \ll (n + m)$) to display.

Each advertiser $i$ submits a $\text{bid}_i$ corresponding to its private click value $v_i$. Given an auction mechanism $\mathcal{M} \langle \mathcal{A}, \mathcal{P} \rangle$ with allocation rule $\mathcal{A}$ and payment rule $\mathcal{P}$, the expected utility $u_i$ for an advertiser $i$ is defined as:
\begin{equation}
    u_i(v_i; \bm{B}) = (v_i - p_i) \cdot \text{ctr}_i,
\end{equation}
where $\bm{B}$ denotes the bidding profile for all advertisers and $p_i$ is the payment of $i$-th advertiser. Two key economic constraints in auction mechanism must be satisfied: incentive compatibility (IC) and individual rationality (IR).
IC requires that truthful bidding maximizes the advertiser's utility. For $\text{ad}_i$, it holds that 
\begin{equation}
    u_i(v_i; v_i, \bm{B}_{-i}) \geq u_i(v_i; \text{bid}_i, \bm{B}_{-i}), \forall v_i, \text{bid}_i \in \mathbb{R}^+.
\end{equation}
IR requires that no advertiser pays more than their bid, that is, $p_i \leq \text{bid}_i, \forall i \in [n]$.

Our objective is maximize the expected platform revenue and order volume with the auction $\mathcal{M}$:
\begin{align}
    \max_{\mathcal{M}}\; \mathbb{E}_{a\in\mathcal{A}}\!\left[\text{Rev}\!+ \!\alpha\!\cdot\!\text{Ord}\right] &=\mathbb{E}_{a\in\mathcal{A}}\! \left( \sum_{i=1}^{k} p_i \!\cdot\!\text{ctr}_i\!+\!\alpha\!\cdot\!\text{ctr}_i\!\cdot\! \text{cvr}_i \!\right),\\
    &\text{s.t. IC and IR constraints},\nonumber
    \label{eqn:obj}
\end{align}
where $\alpha$ is a weight coefficient to balance revenue and orders.

\section{Methodology}
\begin{figure}[ht]
	\centering
	\includegraphics[width=\linewidth,angle=0]{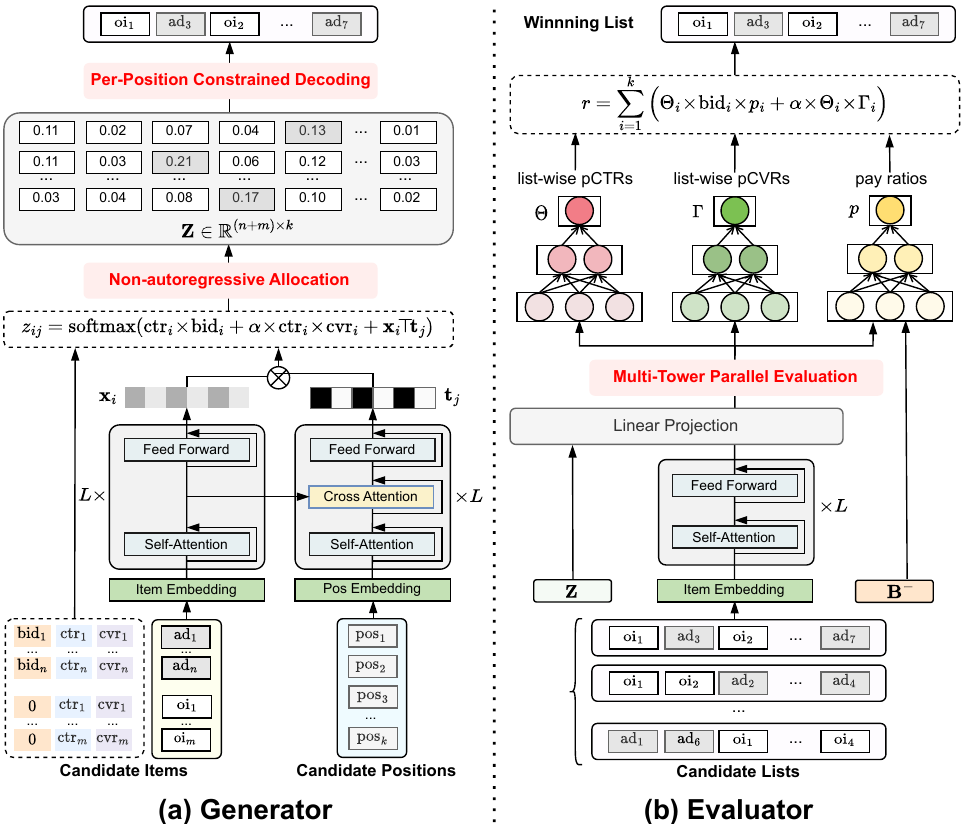}
	\caption{Overview of the NGA architecture.}
	\label{fig:model}
\end{figure}
In this section, we introduce the overall framework of the proposed Non-autoregressive Generative Auction with global externalities (NGA), as illustrated in Figure~\ref{fig:model}. NGA consists of two major components: the \textit{Generator}, which efficiently allocates ads and organic content via non-autoregressive and per-position constrained decoding; and the \textit{Evaluator}, which jointly models list-wise externalities through multi-tower parallel evaluation.

\subsection{Generator}
Given a set of $n$ candidate ads and $m$ organic items, the Generator aims to construct an optimal allocation for $k$ slots to display. Each candidate item $i$ is associated with features vector $\bm{e}_i$. In order to capture both item-level and position-level dependencies, we use a dual encoder architecture.
\subsubsection{Item Encoder}
Each candidate item $i$ is represented as an embedding $\bm{x}_i$ by stacking its feature vector $\bm{e}_i$, which is then processed by $L$ transformer blocks of self-attention and feed-forward networks \cite{vaswani2017attention}:
\begin{equation}
    \bm{x}_i = \text{ItemEncoder}(\bm{e}_i).
\end{equation}
\subsubsection{Position Encoder}
Each candidate position $j$ is embedded as $\bm{t}_j$ using learnable position embeddings, further refined by $L$ blocks of self-attention and cross-attention \cite{ren2024non} to attend over item representations:
\begin{equation}
    \bm{t}_j = \text{PosEncoder}(j;\, \{\bm{x}_i\}_{i=1}^{n+m}).
\end{equation}

\subsubsection{Non-autoregressive Allocation}
The allocation probability matrix $\bm{Z} \in \mathbb{R}^{(n+m)\times k}$, containing per-position scores for all candidate items, is computed as:
\begin{equation}
    z_{ij} = \text{softmax}(\text{ctr}_i \times \text{bid}_i + \alpha \times \text{ctr}_i \times \text{cvr}_i + \bm{x}_i^\top \bm{t}_j),
    \label{eq:alloc_prob}
\end{equation}
where $\bm{x}_i^\top \bm{t}_j$ denotes the hidden score.

\subsubsection{Per-position Constrained Decoding}
To address various business constraints commonly encountered in industrial advertising systems, such as minimum starting position for ads, ad density limitations, and de-duplication across brands, we employ a per-position constrained decoding strategy. Specifically, for each display position $j$, we select the candidate item $i$ with the highest allocation probability $z_{ij}$, subject to business-specific constraints.
Formally, at each position $j$, the decoding process searches for the optimal item according to:
\begin{equation}
i^*_j = \arg\max_{i \in \mathcal{I}_j} \bm{Z}_{ij}
\end{equation}
where $\mathcal{I}_j$ denotes the set of candidate items that satisfy all applicable business constraints for position $j$.

This constraint-aware and position-wise decoding framework enables the generator to produce feasible and high-quality allocation sequences that meet both platform policy and user experience requirements in real time.

\subsection{Evaluator}
% The generated candidate lists are subsequently evaluated by the Evaluator, which consists of three parallel towers responsible for estimating list-wise pCTRs $\bm{\Theta}$, list-wise pCVRs $\bm{\Gamma}$, and payment ratios $\bm{p}$.
Given a set of candidate lists generated by the Generator, the Evaluator is responsible for scoring each list to determine the final winning allocation. To comprehensively model user feedback and auction externalities, the evaluator employs a multi-tower architecture to predict list-wise metrics and compute the overall reward.

\subsubsection{List Encoder}
For each candidate list $\bm{Y} = [y_1, y_2, \ldots, y_k]$, where $k$ is the list length, we construct the list embedding as follows:
\begin{equation}
    \bm{X}_{\text{list}} = \text{ListEncoder}\left(\left[\bm{e}_{y_1}; \bm{e}_{y_2}; \ldots; \bm{e}_{y_k}\right]\right).
\end{equation}
Here, $\bm{e}_{y_i}$ denotes the embedding of item $y_i$ and the transformer block consists of $L$ layers of self-attention and feed-forward networks, enabling the modeling of intra-list dependencies and contextual effects.

\subsubsection{Multi-Tower Parallel Evaluation}
The list embedding $\bm{X}_{\text{list}}$ is passed through linear projection with the allocation probability matrix $\bm{Z}$.
Then we utilize three parallel feed-forward towers to predict three tasks, as shown in Figure \ref{fig:model}:
\begin{align}
    \bm{\Theta} &= \text{Tower}(\bm{X}_{\text{list}}; \bm{Z})\\
    \bm{\Gamma} &= \text{Tower}(\bm{X}_{\text{list}}; \bm{Z})\\
    \bm{p} &= \text{Tower}(\bm{X}_{\text{list}}; \bm{Z}; \bm{B}^-),
\end{align}
where $\text{Tower}(\cdot) = \text{Sigmoid}(\text{MLP}(\cdot))$, 
 $\bm{B}^-=\{\bm{b}_{-1},\bm{b}_{-2},...,\bm{b}_{-k}\}\in \mathbb{R}^{k\times (k-1)}$ is the self-exclusion bidding profile,
$ \bm{\Theta} \in \mathbb{R}^{k \times 1}$ is the list-wise pCTRs, $ \bm{\Gamma} \in \mathbb{R}^{k \times 1}$ is the list-wise pCVRs, $\bm{p} \in \mathbb{R}^{k \times 1}$ is the payment ratios.
The final payment is defined as $\text{pay}_i = \text{bid}_i \times p_i$.

\subsubsection{Overall Reward}
For each candidate list, the evaluator computes the overall reward by aggregating the tower outputs as:
\begin{equation}
    R = \sum_{i=1}^{k} \left(\Theta_i \times \mathrm{bid}_i \times p_i + \alpha \times \Theta_i \times \Gamma_i\right).
\end{equation}
The candidate list with the highest reward $R$ is selected as the winning list for display. Besides, by structuring the evaluator as a multi-tower parallel network, NGA can simultaneously predict list-wise metrics and compute payment, which significantly improves inference efficiency and system scalability compared to traditional serial evaluation methods.

\subsection{Optimization and Training}
\subsubsection{Optimization of Evaluator}
We first train the Evaluator to capture permutation-level externalities using users' real feedback signals, including clicks and conversions. The Evaluator is optimized by minimizing the following loss:
\begin{align}
\mathcal{L}_{\text{pctr}}\! &=\! -\frac{1}{|\mathcal{D}|}\! \sum_{d\in\mathcal{D}}\! \sum_{i=1}^k \left( y_i^{\text{pctr}} \log \Theta_i + (1 - y_i^{\text{pctr}}) \log (1 - \Theta_i) \right)\\
\mathcal{L}_{\text{pcvr}}\! &=\! -\frac{1}{|\mathcal{D}_c|}\! \sum_{d\in\mathcal{D}_c}\! \sum_{i=1}^k\! \left( y_i^{\text{pcvr}} \log \Gamma_i \!+\! (1 - y_i^{\text{pcvr}}) \log (1 - \Gamma_i) \right)\\
\mathcal{L}_{\text{pay}} \! &=\! -\frac{1}{|\mathcal{D}|}\sum_{d \in \mathcal{D}}\! \left(
    \sum_i^k \text{bid}_i p_i \Theta_i \!-\! \sum_i^k \lambda_i \widehat{\text{rgt}}_i \!-\! \frac{\rho}{2}  \sum_i^k (\widehat{\text{rgt}}_i)^2
    \right),
\end{align}
where $y^{\text{pcxr}}$ represents ground-truth labels derived from real user interactions, $\mathcal{D}$ is the exposed dataset and $\mathcal{D}_c$ is the clicked dataset, $\widehat{\text{rgt}}_i$ is the ex-post regret of $y_i$, $p_i$ is the predicted payment ratio, $\lambda_{i}$ is a Lagrange multiplier, and $\rho>0$ is the hyperparameter for the IC penalty term. The final loss of the Evaluator is
\begin{equation}
    \mathcal{L}_{\text{E}} = w_1*\mathcal{L}_{\text{pctr}} + w_2*\mathcal{L}_{\text{pcvr}} + w_3*\mathcal{L}_{\text{pay}},
\end{equation}
where $w_i$ is the hyperparameter weight.

\subsubsection{Optimization of Generator}
After convergence of the Evaluator, we adopt a policy gradient based method to train the Generator. Given a generated winning item sequence $\bm{Y} = \{y_1,y_2,...,y_k\}$, we define the marginal contribution of each item $y_i$ to platform revenue as:
\begin{equation}
r_{y_i} = R_{\bm{Y}} - R_{\bm{Y}^{-i}},
\end{equation}
where $\bm{Y}^{-i}$ denotes the best alternative item sequence excluding $y_i$.
We then apply a policy gradient objective to maximize the expected rewards:
\begin{equation}
\mathcal{L}_{\text{G}} = -\frac{1}{|\mathcal{D}|} \sum_{d\in\mathcal{D}} \sum_{y_i\in \bm{Y}} r_{y_i} \log Z_{y_i},
\end{equation}
where $Z_{y_i}$ is the allocation probability for item $y_i$ by Equation~\eqref{eq:alloc_prob}. This design encourages the generator to produce sequences that yield higher overall revenue, using fixed reward model parameters.

\section{Experiments}
In this section, we systematically evaluate the proposed NGA model on real-world industrial advertising data, focusing on the following research questions:
\begin{itemize}[leftmargin=2em,topsep=2pt]
\item \textbf{RQ1}: How does NGA compare with representative state-of-the-art advertising models in terms of key business metrics?
\item \textbf{RQ2}: What is the practical effectiveness and efficiency of NGA when deployed in online industrial scenarios?
\end{itemize}

\subsection{Experimental Setup}
\subsubsection{Dataset}
We conduct experiments on a large-scale dataset collected from a commercial location-based services (LBS) platform, covering the period from May 6 to May 20, 2025. The dataset contains approximately 225 million page-level user requests, over 60 million distinct users, and nearly 200 million unique candidate items. To ensure a fair and temporally consistent evaluation, we use the first 14 days for model training and reserve the last day for testing.
% 15000000*15 = 22500 0000

\subsubsection{Evaluation Metrics}
We employ the following metrics to comprehensively assess platform revenue, user experience, and advertiser incentive compatibility:

\begin{itemize}[leftmargin=2em, topsep=2pt]
    \item \textbf{RPM (Revenue Per Mille)}: Revenue generated per 1,000 impressions, calculated as \(\text{RPM} = \frac{\sum \text{clicks} \times \text{payment}}{\sum \text{impressions}} \times 1000\).
    
    \item \textbf{CTR}: \(\text{CTR} = \frac{\sum \text{clicks}}{\sum \text{impressions}} \times 100\%\).
    
    \item \textbf{CVR}: \(\text{CVR} = \frac{\sum \text{conversions}}{\sum \text{clicks}} \times 100\%\).
    
    \item \textbf{IC Metric (\(\Psi\))}: Incentive compatibility regret, defined as \(\Psi = \frac{1}{\mathcal{D}} \sum_{d \in \mathcal{D}} \sum_{i \in k} \frac{\widehat{\text{rgt}}_i^d}{u_i(v_i^d, b^d)}\), where \(\widehat{\text{rgt}}_i^d\) is the empirical ex-post regret and \(u_i(v_i^d, b^d)\) denotes the advertiser utility. Lower values indicate better truth-telling incentives~\cite{deng2020data,liao2022nma,wang2022designing}.
\end{itemize}
For offline experiments, all metrics are computed based on predictions from the reward model, and for commercial confidentiality reasons, the absolute value has been converted.
% For online testing, real-world interaction data is used.

\subsubsection{Baselines}
We compare NGA against the following state-of-the-art auctions, and the ad slot allocation for all baselines is fixed position, which has the same ratio of ad impressions as NGA.

\begin{itemize}[leftmargin=2em, topsep=2pt]
    \item \textbf{uGSP~\cite{bachrach2014optimising}}: The classical second-price auction mechanism that ranks ads by ($\text{bid}\!\cdot\!\text{ctr}\!+\!\alpha\!\cdot\!\text{ctr}\!\cdot\! \text{cvr}$), without modeling externalities.
    \item \textbf{DNA~\cite{liu2021neural}}: An end-to-end neural auction framework that learns set-level externalities for ad auction.
    \item \textbf{CGA~\cite{zhu2024contextual}}: A generative auction model that explicitly captures permutation-level externalities for multi-slot ad auction.
\end{itemize}

\subsubsection{Implementation Details}
We use the Adam optimizer~\cite{kingma2014adam} with an initial learning rate of $3 \times 10^{-3}$ and a batch size of 2048. All experiments are conducted on NVIDIA A100 GPUs with 80\,GB memory. We consider a candidate set consisting of $n = 30$ ads and $m = 20$ organic items. The number of allocation slots is set to $k = 10$. Beam search with a beam size of 20 is used during inference to generate multiple candidate sequences. A hyperparameter $\alpha = 5$ is used in both the reward model and the allocation probability matrix to balance revenue and user experience. All experiments were conducted $5$ times and the average results were presented.

\subsection{Offline Performance (RQ1)}
\begin{table}[htbp]
  \centering
  \caption{Offline performance comparison between NGA and baselines on an industrial dataset. Each result is reported as the mean (lift percentage), where lift percentage denotes the improvement of NGA over the best baseline.}
  \label{tab:offline-results}
  \begin{tabular*}{\linewidth}{@{\extracolsep{\fill}}lcccc}
    \toprule
    \textbf{Model} & \textbf{RPM} & \textbf{CTR} & \textbf{CVR} & \textbf{\(\Psi\) (\%)} \\
    \midrule
     uGSP & 200.42\,($-$11.5\%) & 5.26\%\,($-$9.7\%) & 5.10\%\,($-$4.1\%) & 11.4\% \\
    DNA & 210.39\,($-$7.1\%) & 5.54\%\,($-$5.4\%) & 5.14\%\,($-$3.4\%) & 8.5\% \\
    CGA & 221.49\,($-$2.2\%) & 5.75\%\,($-$1.8\%) & 4.98\%\,($-$6.4\%) & 3.8\% \\
    \midrule
    \textbf{NGA} & \textbf{226.47} & \textbf{5.83\%} & \textbf{5.32\%} & \textbf{2.9\%} \\
    \bottomrule
  \end{tabular*}
  \vspace{2pt}
\end{table}

% 生成式增加搜索空间，而不是局限于启发式生成的有限的序列
% NGA相对于CGA有额外引入自然的建模，DNA没有list-wise的外部性建模，uGSP没有外部性建模。  
% 为什么CTR的涨幅比CVR高，因为外部性的建模主要影响点击 对于转化的影响相对小一些

% Table~\ref{tab:offline-results} shows that NGA achieves significant improvements over all baselines across key business metrics. The RPM gain over the best baseline CGA demonstrates NGA’s stronger ability to optimize platform revenue, mainly due to its explicit modeling of both ads and organic content. The CTR increase benefits from incorporating global contextual signals, while the even larger CVR lift reflects the multi-objective design of NGA, since CGA only considers CTR. NGA also achieves the lowest IC regret $\Psi$ because multi-tower joint training improves both model capacity and learning efficiency. Compared with DNA, NGA’s advantage mainly comes from its complete context modeling. Although DNA supports multi-objective optimization, it does not explicitly capture externalities, so NGA achieves a greater RPM and CTR improvement. The results over uGSP further confirm the benefit of advanced context-aware auction design. Overall, NGA demonstrates superior effectiveness, making it highly suitable for deployment in large-scale industrial advertising environments.

Table~\ref{tab:offline-results} shows that NGA achieves significant improvements over all baselines across key business metrics. Specifically, NGA's RPM of 226.47 marks a $2.2\%$ increase over CGA, demonstrating its enhanced ability to optimize platform revenue due to the explicit modeling of both ads and organic content. With a CTR of $5.83\%$, NGA shows a $1.8\%$ improvement, benefiting from global contextual signals that heighten user interaction, while the CVR lift to $5.32\%$ is notably $6.4\%$ higher than CGA, underscoring NGA's multi-objective design. NGA also achieves the lowest IC regret \(\Psi = 2.9\%\), which indicates enhanced incentive compatibility through multi-tower training. Compared with DNA, NGA’s advantage mainly comes from its complete context modeling. Although DNA supports multi-objective optimization, it does not explicitly capture golbal externalities, so NGA achieves a greater RPM and CTR improvement. The results over uGSP further confirm the benefit of advanced context-aware auction design. Overall, NGA demonstrates superior effectiveness, making it highly suitable for deployment in large-scale industrial advertising environments.
% In comparison with DNA, NGA's advantage mainly arises from comprehensive context modeling; despite DNA's support for multi-objective optimization, it lacks explicit externality modeling. Therefore, NGA achieves a greater RPM and CTR improvement, securing its position as highly effective in large-scale industrial advertising environments. Results over uGSP further confirm the benefit of its advanced context-aware auction design.

\subsection{Online Experiments (RQ2)}

\begin{table}[htbp]
  \centering
  \caption{Relative improvements of NGA over multi-objective version of CGA with fixed position in online A/B tests.}
  \label{tab:online-results}
  \small % 或 \footnotesize
  \begin{tabular*}{\linewidth}{@{\extracolsep{\fill}}lcccc}
    \toprule
    \textbf{Relative change in metrics} & \textbf{RPM} & \textbf{CTR} & \textbf{CVR} & \textbf{RT} \\
    \midrule
    NGA over baseline-CGA  & $+1.9\%$ & $+1.6\%$ & $+0.8\%$ & $-6.5\%$ \\
    \bottomrule
  \end{tabular*}
  \vspace{2pt}
\end{table}

% We conduct large-scale online A/B tests on a commercial advertising platform, where NGA is compared with a strong CGA baseline using 2\% of live traffic. As shown in Table~\ref{tab:online-results}, NGA consistently outperforms CGA across key business metrics, including RPM, CTR, and CTCVR. These improvements are attributed to NGA’s explicit modeling of global externalities between ads and organic content, as well as its non-autoregressive and parallel architecture, which ensures real-time efficiency without introducing additional latency. The results demonstrate that NGA is both effective and efficient for industrial online advertising deployment.

The industrial online auction model is a multi-objective version of CGA constructed based on business constraints. Compared with the CGA method in the original paper, it considers both revenue and order quantity objectives.
Table~\ref{tab:online-results} presents the results of online A/B tests conducted on a commercial advertising platform with 2\% of total production traffic, comparing NGA with the strong CGA baseline using fixed ad slots. NGA delivers consistent and significant improvements across all key business metrics. Specifically, NGA achieves a $1.9\%$ increase in RPM, a $1.6\%$ lift in CTR, and a $0.8\%$ gain in CVR compared to CGA. In terms of efficiency, NGA maintains real-time inference with a $6.5\%$ decrease in response time (RT), which remains well within practical limits for industrial deployment. This efficiency gain becomes even more significant as the required number of generated sequences increases, further demonstrating NGA’s suitability for large-scale online advertising systems.

\section{Conclusion}
\label{sec:conc}

In this paper, we presented the Non-autoregressive Generative Auction with global externalities (NGA), a novel end-to-end auction framework for industrial online advertising. NGA advances the state-of-the-art methods by jointly modeling global externalities between ads and organic content, employing a non-autoregressive, constraint-based decoding mechanism and a parallel multi-tower evaluator for list-wise reward and payment computation. Extensive offline experiments and large-scale online A/B tests on a commercial advertising platform demonstrate that NGA consistently outperforms strong baselines in terms of both effectiveness and efficiency, achieving notable improvements in platform revenue, user experience, and incentive compatibility. Looking ahead, our framework offers a promising direction for large-scale, real-time auction design and deployment in online advertising. Future work will extend NGA to support more complex business constraints and generalize the approach to other channel settings.

% \section*{GenAI Disclosure Statement}
% We acknowledge that generative AI was only used for language editing in a manner comparable to basic spelling and grammar correction, consistent with disclosure guidelines.

% \begin{acks}
% The generative tools and 
% \end{acks}

% \newpage
\bibliographystyle{ACM-Reference-Format}
\balance
\bibliography{ref}

\end{document}